\documentclass{IEEEtran}
\usepackage{cite}
\usepackage{amsmath,amssymb,amsfonts}
\usepackage{algorithmic}
\usepackage{graphicx}
\usepackage{textcomp}
\usepackage{bm}
\usepackage{color}
\def\BibTeX{{\rm B\kern-.05em{\sc i\kern-.025em b}\kern-.08em
    T\kern-.1667em\lower.7ex\hbox{E}\kern-.125emX}}
\begin{document}

\title{{\fontsize{24}{26}\selectfont{Communication\rule{29.9pc}{0.5pt}}}\break\fontsize{16}{18}\selectfont
Pseudospin-Polarized Topological Line Defects in Dielectric Photonic Crystals}

\author{Menglin~L.~N.~Chen, \IEEEmembership{Member, IEEE},
        Li~Jun~Jiang, \IEEEmembership{Fellow, IEEE}, Zhihao Lan,
        and~Wei~E.~I.~Sha, \IEEEmembership{Senior Member, IEEE}
\thanks{M. L. N.~Chen and L. J. Jiang are with the Department of Electrical and Electronic Engineering, The University of Hong Kong, Hong Kong (e-mail: menglin@connect.hku.hk; jianglj@hku.hk).}
\thanks{Z. Lan is with the Department of Electronic and Electrical Engineering, University College London, United Kingdom (email: z.lan@ucl.ac.uk).}%
\thanks{W.~E. I.~Sha is with the Key Laboratory of Micro-nano Electronic Devices and Smart Systems of Zhejiang Province, College of Information Science and Electronic Engineering, Zhejiang University, Hangzhou 310027, China and is also with the Department of Electronic and Electrical Engineering, University College London, United Kingdom (email: weisha@zju.edu.cn).}}

\markboth{IEEE transactions on antennas and propagation,~Vol.~XX, No.~XX, August~2019}%
{Shell \MakeLowercase{\textit{et al.}}: Bare Demo of IEEEtran.cls for IEEE Journals}

\maketitle

\begin{abstract}
Electromagnetic topological insulators have been explored extensively due to the robust edge states they support. In this work, we propose a topological electromagnetic system based on a line defect in topologically nontrivial photonic crystals (PCs). With a finite-difference supercell approach, modal analysis of the PCs structure is investigated in detail. The topological line-defect states are pseudospin polarized and their energy flow directions are determined by the corresponding pseudospin helicities. These states can be excited by using two spatially-symmetric line-source arrays carrying orbital angular momenta. The feature of the unidirectional propagation is demonstrated and it is stable when disorders are introduced to the PCs structure.
\end{abstract}


\section{Introduction}

The concept of topology is put forward along with the discovery of the quantum Hall effects and topological insulators in condensed matter~\cite{klitzing1980new,kane2010colloquium,sczhang2011TI}. By putting two materials with different topologies in contact, there exist edge states at the interface~\cite{hatsugai1993chern,sczhang2006general}. Because the topology is stable against disorders of a system, these states propagating in a robust unidirectional way are called topologically protected. Photonic crystals (PCs) are analogues of solid crystals so there is a similarity between the behavior of photons and electrons~\cite{yablonovitch1987inhibited,john1987strong}. In 2008, it was proven that analogous topological effects also exist in PCs, where no quantum but classical electromagnetic (EM) nature applies~\cite{haldane2008possible}.

Recent research work has demonstrated that various photonic systems can have nontrivial topology~\cite{minghui2018photonics}. For example, a gyromagnetic PC under external magnetic fields has nontrivial bulk topology and unidirectional backscattering-immune edge states have been observed experimentally~\cite{Joannopoulos2009observation}. However, as gyromagnetic effect is weak at optical frequencies and not amenable to on-chip integration, topological photonic systems composed of non-gyrotropic materials, such as helical waveguide arrays~\cite{rechtsman2013photonic} or coupled ring resonators~\cite{hafezi2013imaging}, are proposed as alternative promising platforms. Moreover, topological photonic systems that preserve the time-reversal symmetry (TRS) and exploit the concepts such as quantum spin Hall (QSH)~\cite{Khanikaev2013photonic, Ma2015guiding, jwdong2014experimental,huxiao2015scheme,barik2016two,wuying2016pseudo} or quantum valley Hall (QVH)~\cite{Ma2016all,jwdong2019a,Shalaev2019robust} effects have also been proposed. For example, by designing PCs with hexagonal $C_{6v}$ symmetry, the doubly degenerate dipole and quadrupole modes can be used to realize the photonic QSH states~\cite{huxiao2015scheme}, and the pseudospin-momentum locking behavior has been experimentally observed~\cite{yves2017crystalline,hangzh2018visualization}. Besides, homogeneous media, for example, bianisotropic~\cite{hanson2017berry,karimi2018unidirectional} or hyperbolic~\cite{zhuangshuang2015topolgocial} metamaterials, can also have nontrivial topological properties. For all the topological photonic systems with TRS, the edge states are realized at the interface between two PCs with trivial and nontrivial topologies.

In this work, we propose a novel topological waveguide that is constructed by a line defect in only a single two-dimensional (2D) PC with nontrivial topology, which can have important practical advantages. To numerically analyze the properties of the topological waveguide, we develop a simple and effective supercell approach based on the finite-difference (FD) method, from which the band structures can be quickly obtained. The topological line-defect states are identified in the band structure and successfully excited using two spatially-symmetric line-source arrays. The unidirectional-propagation feature of the defect states is verified through full-wave simulations. Furthermore, we find no noticeable backscattering by introducing disorders to the PCs structure.

\section{Modal Analysis}

\subsection{Bulk States}

\begin{figure}[!tb]
\centering
\includegraphics[width=\columnwidth]{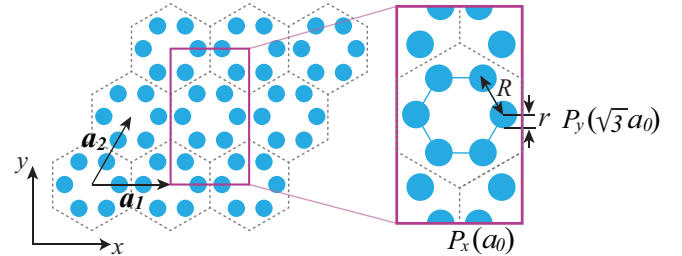}
\caption{Geometry of the 2D PC arranged in a triangular lattice. Each cluster is composed of six cylinders, forming a hexagon with the side length of $R$. The cylinders are pure dielectric, with radius $r$ and dielectric constant $\epsilon$. The right inset shows the smallest supercell holding the periodicity along $x$ and $y$ directions.}
\label{cylinders}
\end{figure}

A 2D PC made using a triangular lattice of hexagonal clusters is shown in Fig.~\ref{cylinders}. $\bm{a_1}$ and $\bm{a_2}$ are the two translation vectors with the length of $a_0$, i.e. the lattice constant. Each cluster is composed of six dielectric cylinders located at the corners of a hexagon, with the side length of $R$.

Only the transverse-magnetic (TM) modes are considered, i.e. electric field only has the out-of-plane component and magnetic field is confined to the $xy$ plane. The governing equation for the TM modes of the 2D PC is written as

\begin{equation}
\frac{1}{\bar{\epsilon}}\frac{\partial^2 E_z}{\partial x^2}+\frac{1}{\bar{\epsilon}}\frac{\partial^2 E_z}{\partial y^2} +k_0^2E_z= 0,
\label{master}
\end{equation}
where $k_0$ is the free-space wave number and $\bar{\epsilon}$ is the averaged dielectric constant~\cite{menglinPRA}.

This equation can be rewritten as $ME_z=k_0^2E_z$ and solved as an eigenvalue problem to obtain eigenvalues $k_0$ and eigenmodes $E_z$. Due to the periodicity of the PC, we can restrict the eigenvalue problem to a single cluster. Nevertheless, to simplify the numerical calculation, we build a rectangular supercell with the periodic boundary conditions imposed on $x$ and $y$ directions. The corresponding Bloch wave numbers are $k_x$ and $k_y$. The lengths of the reassigned translation vectors are $a_0$ along $x$ and $\sqrt{3}a_0$ along $y$. Then, we use the FD method to construct the matrix $M$, which is much easier when compared with integral methods in manipulating the Bloch boundary conditions~\cite{zhengxz2014implementation}. Details on the construction of the matrix $M$ are provided in the Appendix. Then, the eigenvalue problem can be solved. The photonic band structure is drawn by sweeping $k_x$ and $k_y$ along the high symmetry directions of the irreducible Brillouin zone.

\begin{figure*}[!tb]
	\centering
	 \includegraphics[width=2\columnwidth]{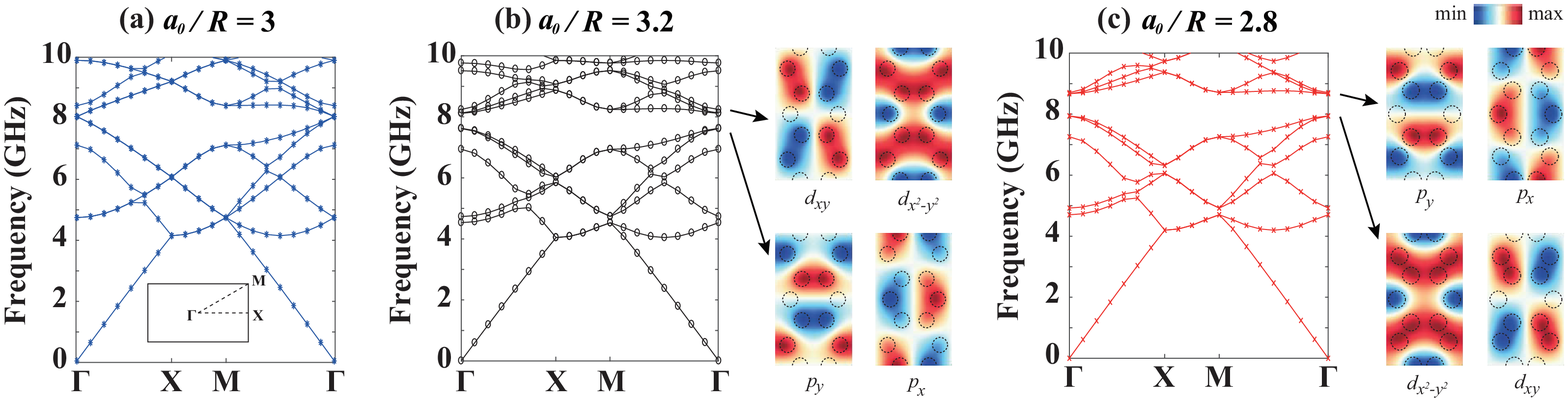}
	\caption{Band structures of the 2D PC in Fig.~\ref{cylinders} and $E_z$ of the dipole and quadrupole states at the $\Gamma$ point in the supercell when (a) $a_0=3R$, (b) $a_0=3.2R$, and (c) $a_0=2.8R$. The dielectric constant of the cylinders $\epsilon=11.7$ and radius $r=2$~mm. The side length of the hexagons $R=6$~mm.}
	\label{BG_bulk}
\end{figure*}

It has been found that the relative sizes of $a_0$ and $3R$ distinguish the topologies of the photonic band structures~\cite{huxiao2015scheme}. We calculate the band structures for three cases with $a_0=2.8R,\,\,3R,\,\,3.2R$. The Brillouin zone is shown in the inset of Fig.~\ref{BG_bulk}(a). When $a_0=3R$, there is no band gap and double Dirac cones with a four-fold degeneracy appear at the $\Gamma$ point. While when $a_0\neq3R$, the four-fold degenerate states at the $\Gamma$ point split into two doubly-degenerate states and the band gap opens. The doubly degenerate states are regarded as dipole ($p_x$, $p_y$), and quadrupole ($d_{xy}$, $d_{x^2-y^2}$) states, because the $E_z$ patterns of the states in hexagonal clusters are isomorphic to $p_x$, $p_y$, $d_{x^2-y^2}$ and $d_{xy}$ electron orbitals. In Fig.~\ref{BG_bulk}(b), i.e. the case of $a_0=3.2R$, the $p_x$ and $p_y$ states are at the $\Gamma$ point of the lower band, and the $d_{xy}$ and $d_{x^2-y^2}$ states are at the $\Gamma$ point of the upper band. However, when $a_0<3R$, there is a band inversion. The $p$ states and $d$ states switch their positions as depicted in Fig.~\ref{BG_bulk}(c). The band inversion between $p$ and $d$ states implies the nontrivial topology of the PC. These results are consistent with the findings in literature, thus validating our supercell approach with the imposed periodic boundary conditions along the $x$ and $y$ directions. The linear combinations of $p_x$ and $p_y$ ($d_{xy}$ and $d_{x^2-y^2}$) provide the up- and down-pseudospin eigenstates that underlie the topological edge states in the PC.

\subsection{Topological Line-defect States}

Topological edge states have been observed at the interface of two PCs with trivial ($a_0>3R$) and nontrivial ($a_0<3R$) topologies~\cite{huxiao2015scheme,hangzh2018visualization} and in the trivial-nontrivial-trivial PC structures~\cite{gao2018unidirectional}. In the following, we will demonstrate topological line-defect states in a topological waveguide which is constructed by introducing an air gap in only one topologically nontrivial PC structure. Although air can be considered as topologically trivial~\cite{silveirinha2015chern}, the air-nontrivial PC interface cannot support edge states because they cannot be confined. By using the nontrivial PC-air-nontrivial PC structure, topological line-defect states with their power concentrated in the PC region are supported.

\begin{figure*}[!tb]
	\centering
	 \includegraphics[width=2\columnwidth]{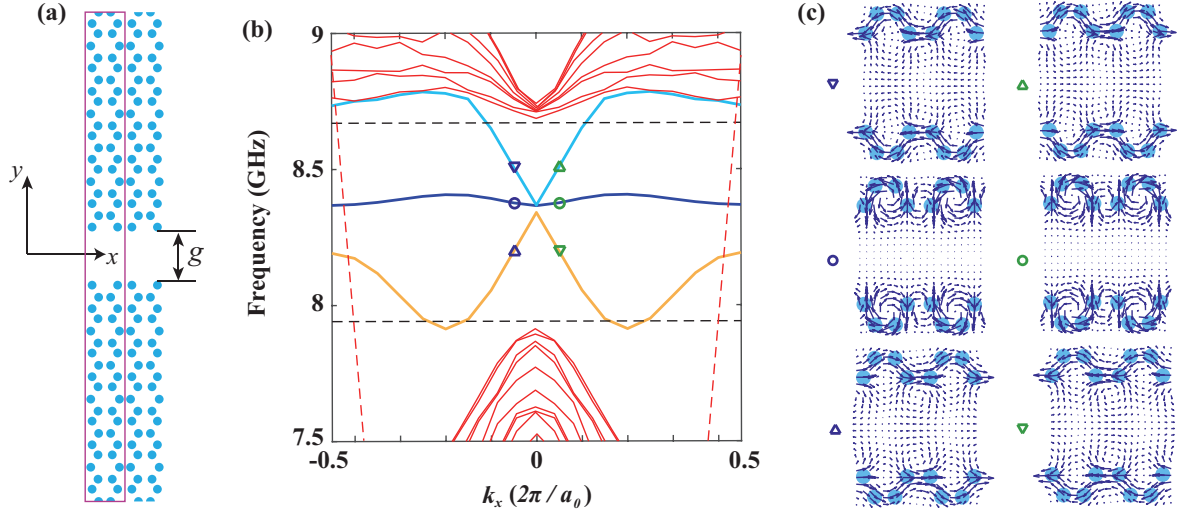}
	\caption{Topological line-defect states. (a) The supercell (indicated by the solid purple rectangle) constructed for the proposed topological waveguide. The width of the gap $g=21$~mm. The parameters of the PC are the same as in Fig.~\ref{BG_bulk}(c). (b) The corresponding band structure. The dashed black lines indicate the upper and lower band edges (same as the band gap calculated in Fig.~\ref{BG_bulk}(c)). The dashed red lines are the light lines $\omega=ck$. Three bands are identified in the band gap. (c) The time-averaged Poynting vector about the line defect at the marked points in (b).}
	\label{FD_Py}
\end{figure*}

The supercell of the proposed waveguide structure is defined in Fig.~\ref{FD_Py}(a) (the solid purple rectangle) with the mirror-symmetry plane, $y=0$. Figure~\ref{FD_Py}(b) illustrates the calculated band structure by sweeping $k_x$ from $-\pi/a_0$ to $\pi/a_0$. The black dashed lines mark the band gap ($7.94$~GHz to $8.67$~GHz) that is calculated in Fig.~\ref{BG_bulk}(c). We find three bands within the band gap. The first band has frequencies lower than $8.34$~GHz and the third band has frequencies larger than $8.37$~GHz. These two bands possess large group velocity around $k_x=0$ and the second band possesses nearly zero group velocity. The distributions of the corresponding time-averaged Poynting vectors at the six marked locations are plotted in Fig.~\ref{FD_Py}(c). Resulting from the structure symmetry, for all the states, the Poynting vectors possess the same mirror-symmetry about $y=0$. However, there is a crucial difference in their energy flow paths. The EM energy of the modes of the first and third bands flow from one supercell to its adjacent supercell. On both the lower and upper edges of the line defect, for the modes marked by triangles, the net energy flows are along right and for the modes marked by the inverted triangles, they are along left. The left- and right-moving paths are accompanied by half-cycle orbits. The rotation of the Poynting vectors along the half-cycle orbits contributes to the net flow of the energy. The direction of rotation correlates with the direction of the energy flow, which implies a pseudospin-locking unidirectional propagation. It is similar to the helical edge states in QSH effects. Meanwhile, we can see from the location of the light lines (the red dashed lines) that these states cannot be guided if they are exposed to air. It is because of the symmetric line defect, the fields at the two edges are coupled and well confined in the PC region. For the modes of the second band, the energy flows are within each supercell, and no effective coupling path is formed between adjacent supercells, which is useless for the guided-wave application.

\begin{figure}[!tb]
	\centering
	 \includegraphics[width=\columnwidth]{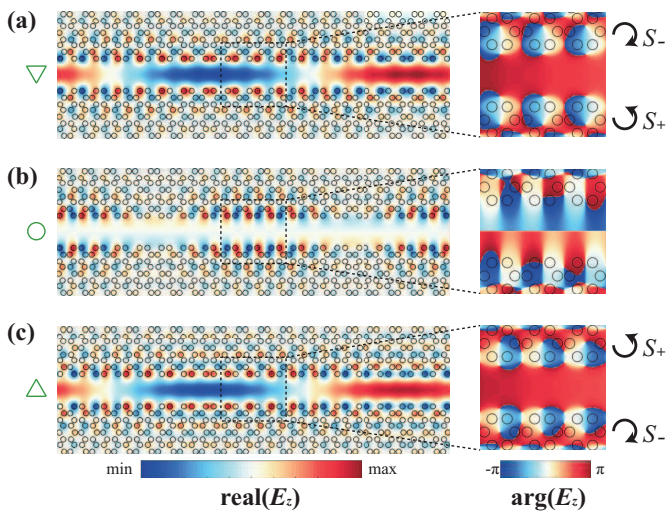}
	\caption{The electric fields (real-space distribution, phase and polarization) at the three marked points with $k_x>0$ on the band structure in Fig.~\ref{FD_Py}(b). (a) The even symmetric mode with lower mode frequency. (b) The odd symmetric mode. (c) The even symmetric mode with higher mode frequency.}
	\label{FD_Efield}
\end{figure}

To further understand the band structure, in Fig.~\ref{FD_Efield}, we plot $E_z$ at the three marked points in Fig.~\ref{FD_Py}(b) with $k_x>0$. Clearly, the modes of the first and third bands have even symmetry, while the modes of the second band have odd symmetry. Based on the previous discussions, the even symmetric modes are topologically protected modes of practical interests. The magnitude of the EM field is strong within the air-gap channel for the even symmetric modes (Figs.~\ref{FD_Efield}(a) and (c)) and is nearly zero for the odd symmetric modes (Fig.~\ref{FD_Efield}(b)). We further examine the phase distributions of these modes. Importantly, it is noted that for the two even symmetric modes, in each hexagon right near the line defect, there is a gradual phase change from $0$ to $2\pi$. The directions of the phase rotation are indicated on the right of Figs.~\ref{FD_Efield}(a) and (c). It can be seen that the EM fields within the hexagons at the upper and lower edges of the line defect are pseudospin polarized with reversed orbital angular momentum (OAM). The OAM of the two even symmetric modes are also opposite.

\section{Excitation of the Topological Waveguide}

\begin{figure}[!tb]
	\centering
	 \includegraphics[width=\columnwidth]{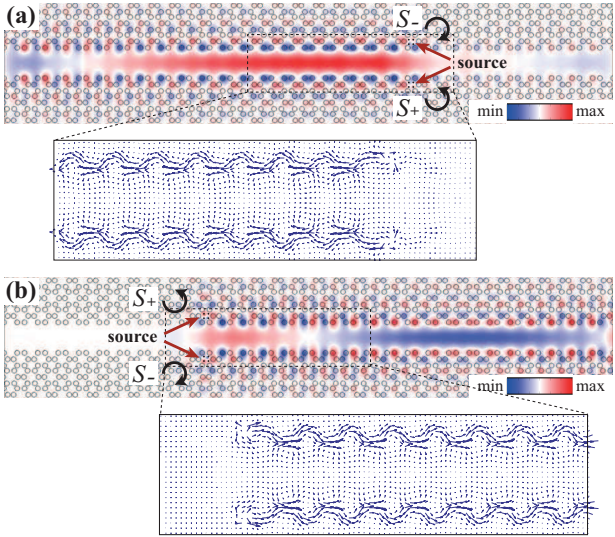}
	\caption{The simulated electric fields (real-space distribution) and the time-averaged Poynting vectors at (a) $f=8.3$~GHz (first band), (b) $f=8.46$~GHz (third band). In each case, there is a pair of sources carrying reversed OAM. The OAM is generated by a four-line-source array in the hexagon near the edge.}
	\label{Ez_sim}
\end{figure}

Based on our analysis, topological line-defect guiding states are the pseudospin-polarized symmetric modes. In the following, we excite these states by using line sources carrying OAM. COMSOL software is employed to simulate the PC structure with scattering boundary conditions enclosing the whole structure. We use a four-line-source array to generate OAM. To match the source symmetry with the eigenstates symmetry, a pair of arrays is put inside the hexagons on the upper and lower edges, which is illustrated in Fig.~\ref{Ez_sim}. A topological line-defect state at the third band in Fig.~\ref{FD_Py}(b) can be selectively excited by setting the signs of OAM of the source array to be the same as in Fig.~\ref{FD_Efield}(c). However, it is worth noting that for the first band, even by selecting the corresponding spin directions, there will be two modes at the frequencies below $8.19$~GHz. The frequency range for the excitation of a pure state is between $8.19$~GHz and $8.34$~GHz.

Simulation results of two excited topological line-defect states of the first and third bands are shown in Fig.~\ref{Ez_sim}. In each case, unidirectional energy propagation is observed and the energy flow to the other direction is suppressed. In Fig.~\ref{Ez_sim}(a), the OAM of the source is set identical to those in the eigenstate in Fig.~\ref{FD_Efield}(a). Therefore, the direction of the energy flow is consistent with the eigenstate in the bottom-right panel in Fig.~\ref{FD_Py}(c), i.e. in both cases, the energy flows leftward. Similarly, Fig.~\ref{Ez_sim}(b) shows the excited topological line-defect state of the third band with the energy moving rightward.

\section{Robustness of the Topological Symmetric Modes}

\begin{figure}[!tb]
	\centering
	\includegraphics[width=\columnwidth]{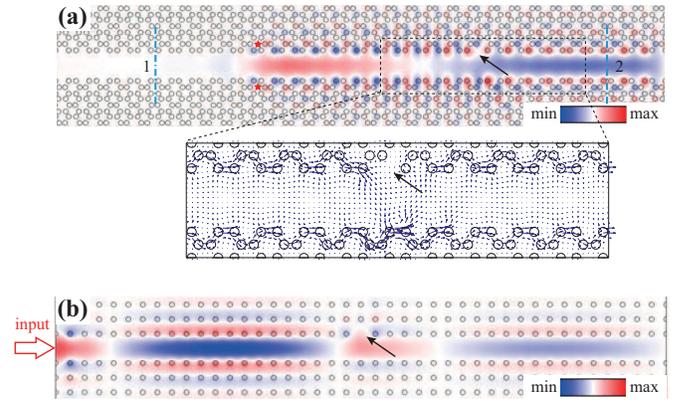}
	\caption{The simulated electric fields (real-space distribution) and the time-averaged Poynting vectors when a cylinder next to the line defect is removed from (a) the proposed topological waveguide at $f=8.46$~GHz and (b) a conventional PC waveguide composed of a square lattice of cylinders at $f=7.95$~GHz. The parameters for the conventional PC: dielectric constant $\epsilon=11.7$, radius $r=2$~mm, and the lattice constant is $12$~mm.}
	\label{Ez_pert}
\end{figure}

The topological line-defect states are immune to bulk diffraction in the presence of defects, which is similar to the waveguide modes in conventional PCs. Beyond that, the topological line-defect states originate from the topology of the PC structure, which makes them more robust when there are disorders. To demonstrate their nature of topological protection, we implement two simulations with the same disorder introduced to the proposed waveguide and a conventional PC waveguide, respectively. In both Figs.~\ref{Ez_pert}(a) and (b), a cylinder on the top edge is removed. Same excitation and operating frequency as in Fig.~\ref{Ez_sim}(b) are used in Fig.~\ref{Ez_pert}(a) and the stars denote the positions of the sources. As can be seen in Fig.~\ref{Ez_pert}(a), the flow of the Poynting vector is distorted around the missing cylinder, but it is reconstructed behind the disorder. The EM energy that passes through the planes $1$ and $2$ (indicated by the dashed blue lines) can be calculated by $U=1/2 \int_l   \text{Re} (\bf{E} \times \bf{H}^*) \cdot \it{d} \mathbf{l}$. Then, we define the backscattering ratio as $U_1/(U_1+U_2)$ and it is calculated to be $1.0\%$. The unidirectional propagation of the EM wave is well maintained in the topological waveguide. As for the conventional PC waveguide in Fig.~\ref{Ez_pert}(b), the simulation frequency is chosen so that it has the same Bloch wave number as the topological waveguide. After removing a cylinder, only $20\%$ power is transmitted and the rest of the power is reflected. Therefore, the topological waveguide is more robust in terms of the unidirectional propagation with disorders than conventional PC waveguide.

\section{Generalized Topological Line-defect States}

The number and frequencies of the line-defect states depend on the width of the air gap. When the gap size decreases, the third band in the band gap will be pushed up to the higher bulk states. When the gap size increases, the first band in the band gap will be pulled down to the lower bulk states. Hence, the line-defect states are different from the edge states holding the bulk-edge correspondence~\cite{graf2013bulk-edge}. At the meantime, the key features of the states keep unchanged, such as the half-cycle orbits of the Poynting vector. However, when the gap becomes infinitely large, the edge states will extend to air and no wave-guiding channel can be formed. The topological waveguide being discussed is symmetric about the $x$ axis, which is a special case. Actually, the air gap can be inserted in a topologically nontrivial PC at other locations, and more generally, the cutting gap can even be on the cylinders.

\begin{figure}[!tb]
	\centering
	\includegraphics[width=\columnwidth]{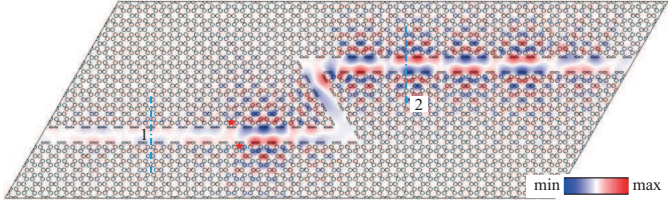}
	\caption{Demonstration of the unidirectional propagation of the topological line-defect state in a bending gap. The plotted frequency is $7.96$~GHz. The width of the air gap is $P_y/2$. The two red stars mark the positions of the two sources carrying OAM.}
	\label{new}
\end{figure}

In Fig.~\ref{new}, a $60$-degree air bend with the width of $P_y/2$ is inserted into the bulk topologically nontrivial PC.  Since the symmetry of the structure changes, the line-defect states being supported now become asymmetric. But the difference in the symmetry property will not change the key features of the topological line-defect states. The new line-defect states still have non-zero group velocity with the inter-supercell energy transfer. At the meantime, the Poynting vector rotates along the half-cycle orbits in the PC on the edge. Importantly, we can observe the unidirectional propagation of this state against sharp bends. The backscattering ratio which is calculated based on the energy transmitted through planes $1$ and $2$ is $8.3\%$.
 
\section{Conclusion}

In summary, we proposed a topological waveguide that supports pseudospin-polarized propagating modes. Unlike any existing designs using two PCs with trivial and nontrivial topologies, our design only contains topologically nontrivial PC with an inserted air gap. The topological line-defect states are supported in the structure, resulting from the coupling between the edge states on upper and lower edges of the line defect. The FD supercell approach was established to calculate the band structures and analyze the field polarization and phase. Moreover, the topological line-defect states are successfully excited by using a pair of sources possessing the same symmetry as the eigenfields and they are found to be immune to disorders in the structure. The demonstration is in microwave but can be scaled up to optical regime.

\appendices
\section{Eigenvalue analysis using FD method}

\begin{figure}[!tb]
	\centering
	\includegraphics[width=\columnwidth]{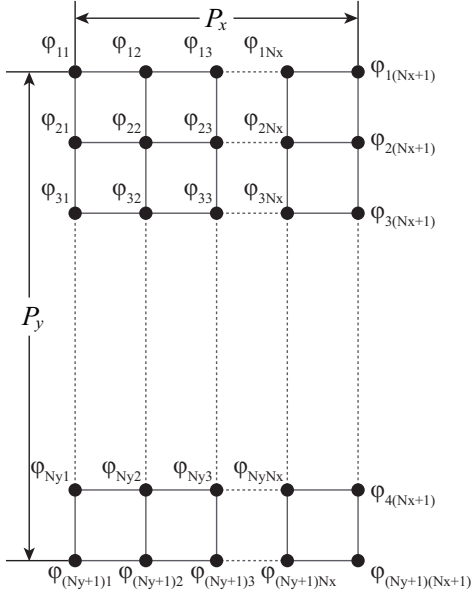}
	\caption{The supercell with FD grids.}
	\label{SI}
\end{figure}

First, the supercell in Fig.~\ref{cylinders} is divided into many grids as shown in Fig.~\ref{SI}. $\phi_{m,n}$ ($m= 1,2,3,...N_y,N_y+1; n=1,2,3,...N_x,N_x+1$) denotes the electric field $E_z$ at each sampling point. Due to the periodicity, each supercell has $N_x\times N_y$ unknowns, i.e. $\phi_{m,n}, m= 1,2,3,...N_y; n=1,2,3,...N_x$. 

To simulate curved boundaries of the dielectric cylinders, considering the tangential $E_z$ component is continuous across the air-dielectric interface, {we calculate the permittivity at each grid point by computing the average of the permittivity at its four surrounding points that are half-grid away along both the $x$ and $y$ directions:

\begin{equation}
\begin{aligned}
\bar \epsilon = \epsilon_{m,n}=& \frac{1}{4}(\epsilon_{m-0.5,n-0.5}+\epsilon_{m+0.5,n-0.5}\\
&+\epsilon_{m-0.5,n+0.5}+\epsilon_{m+0.5,n+0.5}).
\end{aligned}
\end{equation}

Then, the differential eigenvalue equation~\eqref{master} at the grid point, $(m,n)$ is rewritten by using the FD approximation:

\begin{equation}
\begin{aligned}
&\frac{1}{\epsilon_{m,n}}\frac{\phi_{m,n+1}+\phi_{m,n-1}-2\phi_{m,n}}{\Delta x^2}+\\
&\frac{1}{\epsilon_{m,n}}\frac{\phi_{m+1,n}+\phi_{m-1,n}-2\phi_{m,n}}{\Delta y^2}=k_0^2\phi_{m,n}.
\end{aligned}
\label{FD}
\end{equation}

For the grid points going outside of the unknowns, they are treated by the Bloch boundary conditions, i.e.
\begin{equation}
\phi_{m,n}=\phi_{m,n \pm N_x}e^{j \mp k_xP_x}, \quad \phi_{m,n}=\phi_{m \pm N_y,n}e^{j \mp k_yP_y},
\end{equation}
where $k_x$ and $k_y$ are the Bloch wave numbers.

Finally, the differential eigenvalue equation is recast into a matrix form,
\begin{equation}
\begin{aligned}
M\Phi=k_0^2\Phi, \quad \Phi=&(\phi_{11} ~\phi_{12} ~...~ \phi_{1N_x}~ \phi_{21} ~\phi_{22} ~...~ \phi_{2N_x}~...\\
&~\phi_{N_y1} ~\phi_{N_y2} ~...~ \phi_{N_yN_x})^T,
\end{aligned}
\end{equation}
where $M$ is a sparse matrix. The above can be solved by a standard eigenvalue solver in MATLAB.


\section*{Acknowledgement}
This work was supported in part by the Research Grants Council of Hong Kong GRF 17209918, AOARD FA2386-17-1-0010, NSFC 61271158, HKU Seed Fund 201711159228, and Thousand Talents Program for Distinguished Young Scholars of China.

\bibliographystyle{IEEEtran}
\bibliography{reference}

\end{document}